\documentclass[sigconf, nonacm]{acmart}
\AtBeginDocument{%
  }

\setcopyright{acmlicensed}
\copyrightyear{2018}
\acmYear{2018}
\acmDOI{XXXXXXX.XXXXXXX}
\acmConference[Conference acronym 'XX]{Make sure to enter the correct
  conference title from your rights confirmation email}{June 03--05,
  2018}{Woodstock, NY}
\acmISBN{978-1-4503-XXXX-X/2018/06}

\usepackage{subcaption}
\usepackage{tikz}
\usepackage{threeparttable}

\usepackage{setspace}
\usepackage{enumitem}
\setlist[itemize]{noitemsep, topsep=0pt, leftmargin=*}

\DeclareRobustCommand\encircle[1]{\tikz[baseline=(char.base)]{\node[shape=circle,fill,inner sep=0.75pt] (char) {\textcolor{white}{#1}}}}
\begin{document}

%%
%% The "title" command has an optional parameter,
%% allowing the author to define a "short title" to be used in page headers.
\title{PALUTE: Processing-In-Memory Acceleration via Lookup Table for Edge LLM Inference}

%%
%% The "author" command and its associated commands are used to define
%% the authors and their affiliations.
%% Of note is the shared affiliation of the first two authors, and the
%% "authornote" and "authornotemark" commands
%% used to denote shared contribution to the research.

\author{Runyang Tian}
% \authornote{Both authors contributed equally to this research.}
\email{r3tian@ucsd.edu}
\orcid{0009-0000-7298-502X}
% \author{G.K.M. Tobin}
% \authornotemark[1]
% \email{webmaster@marysville-ohio.com}
\affiliation{%
  \institution{University of California San Diego}
  \city{La Jolla}
  \state{CA}
  \country{USA}
}

\author{Yanru Chen}
\email{yac054@ucsd.edu}
\orcid{0009-0006-7958-8725}
\affiliation{%
  \institution{University of California San Diego}
  \city{La Jolla}
  \state{CA}
  \country{USA}}

\author{Weihong Xu}
\email{weihong.xu@epfl.ch}
\orcid{0000-0003-3766-3353}
\affiliation{%
  \institution{Ecole Polytechnique Fédérale de Lausanne}
  \city{Lausanne}
  \country{Switzerland}
}

\author{Tajana Šimunić Rosing}
\email{tajana@ucsd.edu}
\orcid{0000-0002-6954-997X}
\affiliation{%
 \institution{University of California San Diego}
 \city{La Jolla}
 \state{CA}
 \country{USA}}

%%
%% By default, the full list of authors will be used in the page
%% headers. Often, this list is too long, and will overlap
%% other information printed in the page headers. This command allows
%% the author to define a more concise list
%% of authors' names for this purpose.
% \renewcommand{\shortauthors}{Tian et al.}
\renewcommand{\shortauthors}{Tian et al.}

%%
%% The abstract is a short summary of the work to be presented in the
%% article.
\begin{abstract}
Large language models are increasingly deployed on edge devices with tight power and area budgets. While mixed-precision GEMM reduces arithmetic complexity, quantized inference is often dominated by dequantization and nonlinear operators. Lookup Table (LUT)-based method mitigates these costs by precomputing outputs and replacing repeated arithmetic with table lookups, but existing designs incur significant capacity and lookup-latency overheads. This paper presents PALUTE, a LUT-based Processing-In-Memory accelerator built on Monolithic 3D DRAM for efficient edge LLM inference. PALUTE enables in-DRAM LUT queries that exploit the vertical organization of M3D DRAM memory array tiles to achieve high parallelism with low area overhead. A near-memory LUT generator supports low-latency LUT generation for both GEMM and element-wise unary nonlinear operators, while a system-level tiering and scheduling strategy minimizes data movement across memory tiers. Evaluation using cycle-accurate simulation and RTL synthesis shows that PALUTE achieves 1,264 TPS end-to-end throughput at 0.16 W, improving energy efficiency by 12.8$\times$ over CHIME~\cite{CHIME} and 1.6$\times$ over FIGLUT~\cite{FIGLUT}, improving area efficiency by 2.0$\times$ over PIMPAL~\cite{PIMPAL} under W4A4 across Qwen3-4B models.
\end{abstract}

%%
%% The code below is generated by the tool at http://dl.acm.org/ccs.cfm.
%% Please copy and paste the code instead of the example below.
%%


%%
%% Keywords. The author(s) should pick words that accurately describe
%% the work being presented. Separate the keywords with commas.
%% A "teaser" image appears between the author and affiliation
%% information and the body of the document, and typically spans the
%% page.
% \begin{teaserfigure}
%   \includegraphics[width=\textwidth]{sampleteaser}
%   \caption{Seattle Mariners at Spring Training, 2010.}
%   \Description{Enjoying the baseball game from the third-base
%   seats. Ichiro Suzuki preparing to bat.}
%   \label{fig:teaser}
% \end{teaserfigure}

% \received{20 February 2007}
% \received[revised]{12 March 2009}
% \received[accepted]{5 June 2009}

%%
%% This command processes the author and affiliation and title
%% information and builds the first part of the formatted document.
\maketitle

\section{Introduction}
Large language models (LLMs) have emerged as a milestone in Artificial Intelligence (AI), enabling natural and context-aware dialogue. Modern LLMs encode knowledge in billions of parameters, and their scale continues to grow rapidly~\cite{LLM_Survey}. These models support powerful applications such as personalized healthcare agents~\cite{Healthcare_Agent} and interactive intelligent systems~\cite{Intelligent_systems}. However, their increasing size introduces substantial computational and memory demands during inference. Privacy-critical domains such as healthcare further require local execution, accelerating the shift from cloud-centric processing to edge deployment. Edge platforms, however, are constrained by limited compute capability, restricted memory bandwidth, and tight power and area budgets. These constraints are especially severe for decoder-only Transformers, where auto-regressive generation repeatedly performs GEMM and maintains key-value (KV) caches~\cite{Mobile_Edge_Survey}.

Mixed-precision has therefore become a key technique for reducing model size and compute cost by lowering numerical precision~\cite{Quantization_Survey}. Yet many practical systems suffer from substantial dequantization overhead, which can account for 20--90\% of runtime and diminish the benefits of low-precision arithmetic~\cite{QServe}. Although pruning, quantization, and conventional accelerators improve efficiency~\cite{QServe, Atom, QuaRot, I-BERT}, they still preserve the separation between memory and compute. Empirical studies show that data movement dominates system energy, accounting for 62.7\% of total energy on average~\cite{Data_Movement_Bottlenecks}. Thus, reducing arithmetic cost alone cannot resolve the memory bottleneck.

Processing-In-Memory (PIM) addresses this bottleneck by performing computation where data resides, minimizing off-chip transfers~\cite{PIM_Survey}. For example, pLUTo~\cite{pLUTo} demonstrates the potential of DRAM-resident lookup-table (LUT) computation, outperforming CPU and GPU baselines through highly parallel in-memory LUT queries. LUT-based inference also intrinsically avoids dequantization~\cite{FIGLUT}, making it well suited for low-precision LLM workloads.

Prior LUT-based systems~\cite{TMAC, PIMPAL, CHIME, Stratum, FIGLUT} mainly place MAC or LUT units near memory arrays or on logic dies. T-MAC~\cite{TMAC} presents a CPU-friendly LUT-based mpGEMM kernel for low-bit LLM inference, reporting up to $4\times$ higher throughput and 70\% lower energy than llama.cpp. PIMPAL~\cite{PIMPAL} uses parallel in-DRAM arithmetic lookups with locality-aware mapping and LUT aggregation for GEMM, achieving up to speedup of 17.8$\times$ than prior LUT-based PIM designs while reducing area overhead by 40\% over processing unit (PU)-based PIM. To further improve parallelism while reducing power and area, we introduce an energy-efficient, high-throughput PIM architecture for LLM inference using Monolithic 3D (M3D) DRAM, which provides high density and vertical memory array tiers (MATs), enabling vertical LUT queries with high parallelism and low area overhead. Our main contributions are:

    \begin{itemize}
        % \item A detailed architectural analysis and comparison of NMP/PIM compute paradigms, including MAC-based and LUT-based designs, revealing key performance, bandwidth, and energy trade-offs.
        
        \item An LUT query mechanism on M3D DRAM that exploits vertical MATs to achieve high parallelism with low area overhead for dequantization-free low-precision computation.
        
        \item Near-memory LUT generators implemented on the logic die and connected via hybrid bonding, providing low-latency, high-bandwidth support for in-memory LUT querying.

        \item A system-level data tiering, scheduling workflow in M3D DRAM, which minimizes off-chip data movement and alleviates memory bottlenecks during transformer inference.

        % \item A Progressive Token Elimination Unit (PTEU) that adaptively identifies and removes insignificant KV-cache entries, substantially reducing memory footprint and computation cost.
        
        \item  Simulation shows that PALUTE delivers 1,264 TPS at 0.16 W, improving energy efficiency by 12.8$\times$ over CHIME~\cite{CHIME} and 1.6$\times$ over FIGLUT~\cite{FIGLUT}, and area efficiency by 2.0$\times$ over PIMPAL~\cite{PIMPAL}.
    \end{itemize}

\section{Background and Motivations}
     \subsection{Transformer Model}
        
        \begin{figure}[ht]
            \centering
            \includegraphics[width=1\linewidth]{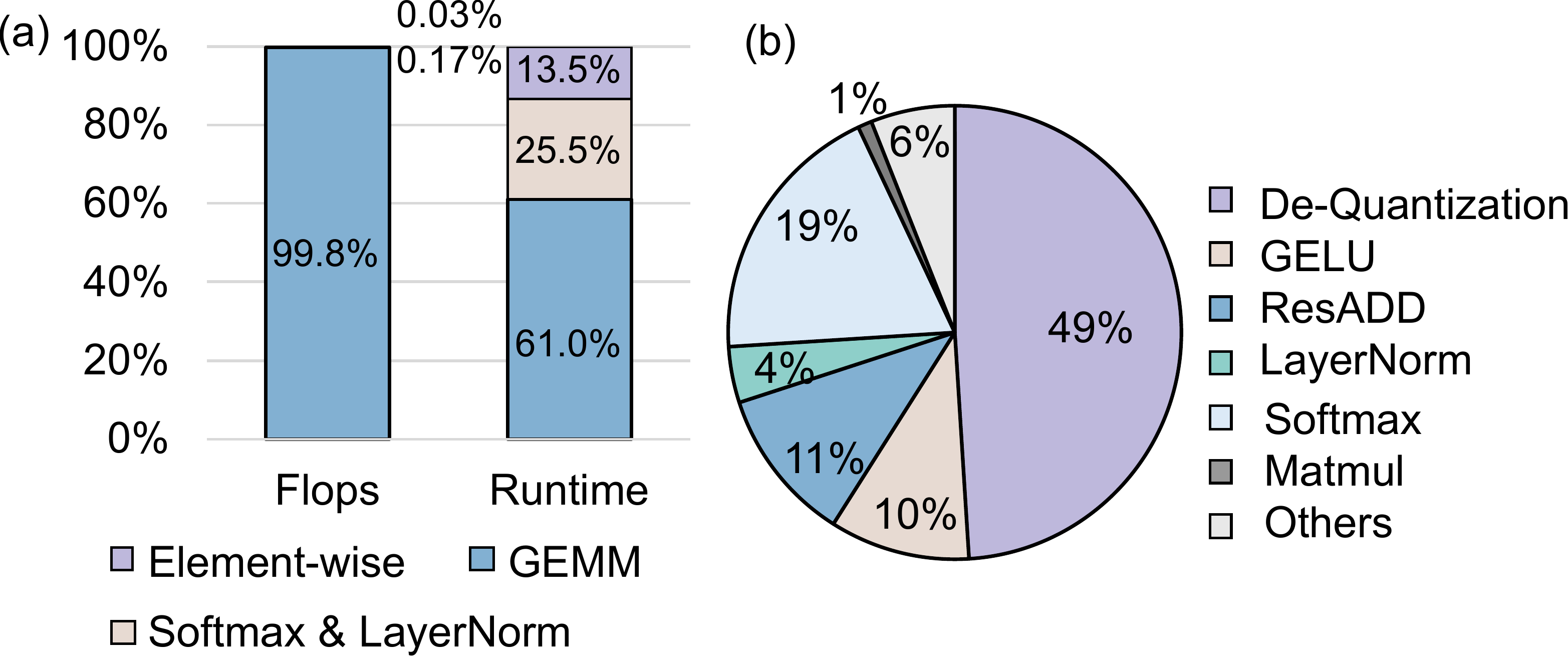}
            \caption{LLM architecture overview (a) Flops and runtime breakdown (b) Runtime breakdown with conventional NPU}
            \label{fig:LLM}
        \end{figure}

        LLM inference is dominated by GEMM and nonlinear operators (e.g., Softmax, GELU). During auto-regressive decoding, these kernels execute once per generated token, so their efficiency directly determines end-to-end latency and energy. Figure~\ref{fig:LLM}(a) shows a clear imbalance: GEMM contributes 99.8\% of FLOPs but only 61.0\% of runtime, while Softmax/LayerNorm/element-wise kernels contribute$<0.2\%$ FLOPs yet consume $\sim$39\% runtime~\cite{DATA_MOVEMENT_IS_ALL_YOU_NEED}. This is a memory-wall effect. These kernels are dominated by data movement between HBM and on-chip SRAM rather than compute~\cite{DATA_MOVEMENT_IS_ALL_YOU_NEED,FLASHATTENTION}. Therefore, optimizing GEMM alone shifts the bottleneck to nonlinear operators and dequantization overhead (Figure~\ref{fig:LLM}(b))~\cite{Transformer_Breakdown}.

    \subsection{In/Near-Memory Processing using Monolithic 3D DRAM}

    The deployment of LLMs is fundamentally constrained by von Neumann bottleneck, the gap between computation capability and memory bandwidth. Limited memory bandwidth makes data movement a dominant source of latency and energy~\cite{DATA_MOVEMENT_IS_ALL_YOU_NEED}. Memory-centric architectures mitigate this bottleneck by co-locating compute with memory, including both near-memory processing (NMP) and Processing-In-Memory (PIM): NMP integrates logic adjacent to memory arrays to exploit high internal bandwidth, while PIM pushes computation closer to or into the memory using device-level primitives for massive parallelism~\cite{Enabling_In-Memory_Computation,DATA_MOVEMENT_IS_ALL_YOU_NEED}. 

\begin{figure}[ht]
        \centering
        \includegraphics[width=1\linewidth]{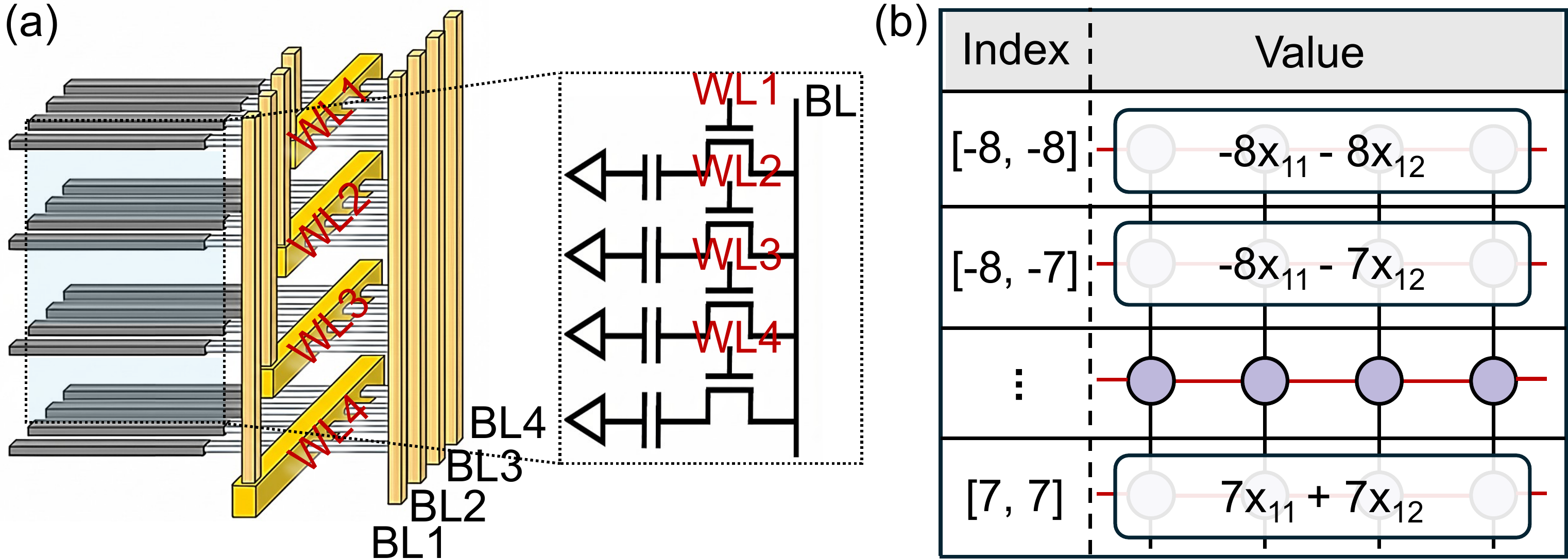}
        \caption{M3D DRAM and LUT structure (a) Vertical M3D DRAM stacking~\cite{M3DDRAM, CHIME} (b) A LUT mapped in M3D DRAM cells}
        \label{fig:M3D_LUT}
    \end{figure}

    M3D DRAM is a key memory substrate for PIM/NMP designs. By sequentially stacking multiple memory tiers on a single wafer and connecting them with dense monolithic inter-tier vias (MIVs), M3D DRAM increases bit density and reduces interconnect latency. Compared with conventional 2D DRAM that relies on long lateral wires or Through-Silicon Via (TSV)-based stacks with relatively sparse vertical connections, M3D DRAM offers denser vertical connectivity with Monolithic Inter-Layer Vias (MIV) and much shorter tier-to-tier interconnect distances. M3D DRAM organizations are commonly categorized into Vertical Wordline (VWL) and Vertical Bitline (VBL) topologies~\cite{M3DDRAM} as shown in Figure~\ref{fig:M3D_LUT}(a). PALUTE adopts a VBL-style organization where bitlines traverse tiers vertically while wordlines are routed within tiers, therefore the DRAM MATs are also vertical, which provides high vertical connectivity that can be exploited for in/near-memory compute and data movement reduction. VBL M3D DRAM can offer higher density (e.g., $\sim$2.74~Gb/mm$^2$) and lower read energy than VWL alternatives, making it attractive for next-generation AI accelerators~\cite{M3DDRAM}.

    \subsection{LUT-based Operation}

    A LUT stores a precomputed mapping from discrete inputs to outputs. At runtime, an input serves as the index, and the corresponding result is returned through a table read. As shown in Figure~\ref{fig:M3D_LUT}(b), each WL represents one index, while cells on that WL store the associated value, which is read out upon WL activation. LUTs trade computation for memory, making them attractive for Transformer inference to reduce redundant MACs in GEMM~\cite{FIGLUT} and accelerate expensive nonlinear operations. They are most effective with low-precision inputs, repeated function invocations, and memory substrates that support highly parallel queries. Their main drawbacks are storage overhead and limited flexibility: LUT size grows rapidly with input precision, and changes in function, range, or precision typically require table regeneration.

    % LUT-based operation can be applied to GEMM and non-linear operations. Two matrix $X\cdot W$ can be split into multiple segments, like $x_{seg1}\cdot w_{seg1}=[x_{1,1},x_{1,2}]\cdot [w_{1,1};w_{2,1}]$. $w_{seg1}$ has fixed possible combinations like $[-8, -8], [-8, -7], \dots, [7,7]$. We can use two elements of $x_{seg1}$ to generate a list of all possible combination of the elements of $x_{seg1}$ and $w_{seg1}$ (table: $-8x_{1,1}-8x_{1,2}, -8x_{1,1}-7x_{1,2}, \dots, 7x_{1,1}+7x_{1,2})$. In this scenario, $w_{seg1}$ becomes index, can find the relative value from the table.

    % For nonlinear operations, element-wise unary operators like GELU, ReLU can be computed by LUT. We can prepare all possible output $GELU(x_{ij}), x_{ij}=-8, -7, \dots 7$, use $x_{ij}$ becomes the index and we can find the result from the table.

    LUT-based computation can be applied to both GEMM and element-wise nonlinear operators. For GEMM, we split $X\cdot W$ into small dot-product segments, e.g., $x_{\mathrm{seg1}}\cdot w_{\mathrm{seg1}}=[x_{1,1},x_{1,2}]\cdot [w_{1,1};w_{2,1}]$.
    When weights are low-bit quantized, $w_{\mathrm{seg1}}$ only takes values from a finite set (e.g., $w_{1,1},w_{2,1}\in[-8,7]$), so the number of possible $w_{\mathrm{seg1}}$ combinations is limited.
    Given $x_{\mathrm{seg1}}$, we precompute a LUT that enumerates all possible results of $x_{\mathrm{seg1}}\cdot w_{\mathrm{seg1}}$, such as $-8x_{1,1}-8x_{1,2}, -8x_{1,1}-7x_{1,2}, \ldots, 7x_{1,1}+7x_{1,2}$.
    At runtime, $w_{\mathrm{seg1}}$ serves as the LUT index and the corresponding partial dot-product value is fetched directly. For nonlinear operators, we use LUTs for element-wise unary functions such as GELU and ReLU. We precompute all possible outputs $f(x_{ij})$ for $x_{ij}\in[-8,7]$ and use $x_{ij}$ as the LUT index to retrieve $f(x_{ij})$.

    \begin{figure*}[t]
        \centering
        \includegraphics[width=1\linewidth]{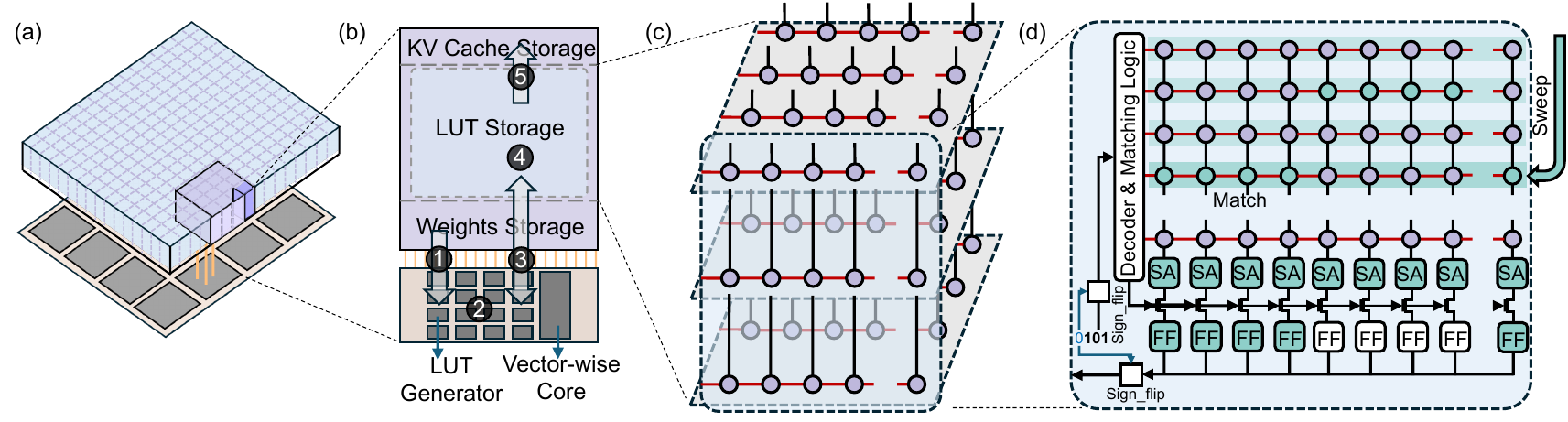}
        \caption{PALUTE hardware design (a) M3D DRAM with logic die (b) Logic die organization, data placement, and scheduling (c) M3D DRAM bank organization with VBLs (d) Vertical M3D MAT with in-DRAM query logic}
        \label{fig:PALUTE_total}
    \end{figure*}
    
    \subsection{Motivation}
    
    LUT-based accelerators replace expensive arithmetic with table reads~\cite{FIGLUT,TMAC}, but LUTs require large capacity and often incur irregular accesses, amplifying memory traffic and single-query latency. As representative SOTA, T-MAC is constrained by CPU memory bandwidth for LUT-driven GEMM~\cite{TMAC}, while FIGLUT, as a standalone ASIC, does not fundamentally resolve the memory wall at the system level~\cite{FIGLUT}. LUT-based PIM/NMP solutions also have limits: pLUTo proposes in-memory (normal DRAM) LUT querying, yet leaves open a complete end-to-end pipeline (e.g., LUT generation/management) and does not fully address LUT-induced latency/area overheads under scaling~\cite{pLUTo}; PIMPAL executes LUTs inside normal DRAM to reduce off-chip transfers, but achievable parallelism and query latency remain bounded by DRAM row-buffer/burst interfaces and LUT footprint, making throughput sensitive to memory-interface limits and access locality~\cite{PIMPAL}.

    To address these challenges, PALUTE’s core innovations are \textbf{(1) a LUT query mechanism using M3D DRAM} that exploits vertical MATs to deliver high parallelism with low area overhead, enabling dequantization-free low-precision computation while accommodating large LUT capacity; \textbf{(2) Near-memory LUT generators on the logic die}, connected via hybrid bonding, that provides low-latency, high-bandwidth support for data movement; and \textbf{(3) a system-level data tiering and scheduling workflow in M3D memory organization} that minimizes off-chip data movement and alleviates memory bottlenecks during Transformer inference.

\section{PALUTE Architecture Design}
    \subsection{System Overview}

        In this section, we present PALUTE, a PIM accelerator that enables LUT-centric LLM inference via in-DRAM LUT querying on M3D DRAM. PALUTE uses an M3D stack with a bottom logic die hybrid-bonded to DRAM tiers (Figure~\ref{fig:PALUTE_total}(a)), exposing high die-to-die bandwidth ($\sim$1.38 TB/s) for near-memory access~\cite{HybridBonding}. We design an in-DRAM LUT querying architecture using M3D DRAM for LLM inference. We decompose GEMM into LUT-mappable partial-MAC combinations, and map these queries onto DRAM MATs. This choice exploits massive MAT-level parallelism and reduces data movement and logic-die compute overhead. (Figure~\ref{fig:PALUTE_total}(d)); and element-wise unary operators (e.g., GELU) are preprocessed into LUTs using the mapping method in Figure~\ref{fig:M3D_LUT}(b). At the system level, PALUTE integrates accumulators into DRAM banks and applies a GEMM storage optimization that uses only a half-table (Sec.~\ref{sec:Hardware}). All the results of LUT are preprocessed in the LUT generator on the logic die (Sec.~\ref{Sec:Hardware}) connected via hybrid bonding.

        PALUTE further provides a scheduling flow that coordinates LUT generation, data placement, and in-DRAM LUT query (Figure~\ref{fig:PALUTE_total}(b)) to minimize cross-tier data movement (Sec.~\ref{Sec:Scheduling}). Finally, M3D DRAM’s vertical MAT organization (VBL) offers high intrinsic parallelism: matching it in planar DRAM would require large area overhead, while TSV-based 3D stacks introduce non-trivial cross-layer latency. PALUTE includes LUT generators on the logic die, which can produce a ready-to-query LUT in three clock cycles, keeping table preparation off the critical path.

    \subsection{PALUTE Hardware Design}\label{sec:Hardware}
            
\begin{figure*}[ht]
                \centering
                \includegraphics[width=1\linewidth]{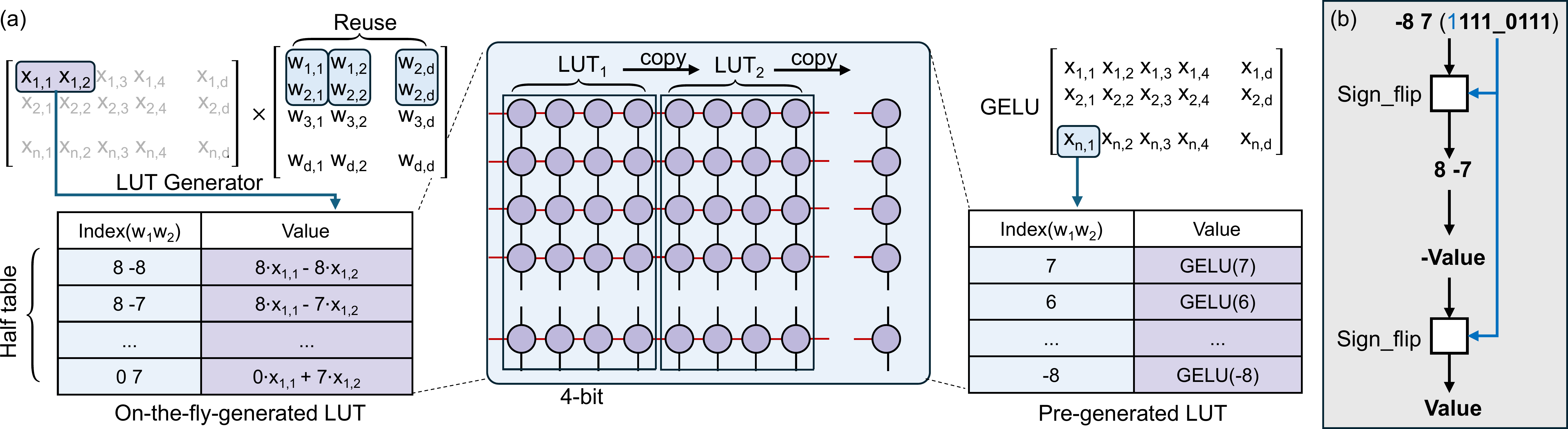}
                \caption{In-DRAM LUT mapping (a) Examples of LUT for GEMM (X$\cdot$W) and GELU (b) Half-table logic for GEMM}
                \label{fig:LUT_workflow}
            \end{figure*}

         LUT-querying supports both GEMM and nonlinear operations. Figure~\ref{fig:LUT_workflow}(a) shows an example where the first row of $X$ multiplies the first row of $W$ and is accumulated; we partition the long dot-product into smaller segments to improve efficiency. The key observation is reuse: a fixed activation segment is shared. Once a segment of $X$ is fixed, it pairs with various segments of $W$, yielding a bounded set of partial sums. Thus, for each activation segment, we precompute all possible results induced by the corresponding weight patterns and store them in a LUT. It encodes each weight segment as an index and directly retrieves the precomputed value. 

        Assuming 4-bit values, LUT entries are stored horizontally across four DRAM cells in an M3D DRAM MAT. We place multiple horizontal replicas of the LUT to increase parallelism, enabling row-wise retrieval of many LUT values simultaneously. Figure~\ref{fig:LUT_workflow}(b) illustrates half-table lookup: due to sign symmetry, we store only entries for non-negative values. Negative cases flip sign bits to map to the symmetric index, read the LUT value, and restore the sign. Unlike FIGLUT’s half flip-flop based LUT (hFFLUT) that reduces flip-flop LUT cost inside an ASIC~\cite{FIGLUT}, PALUTE applies half-table directly to in-DRAM LUT storage to reduce off-chip transfers. This halves LUT capacity demand, reducing DRAM storage overhead and allowing longer LUTs to fit in-memory query. Let segment length = $b$ and weight bit width = $q$, the length of LUT $L=2^{bq}$, PALUTE chooses segment length to be 2.

          \subsubsection{LUT-Query-Supported M3D DRAM MAT}

             To enable highly parallel LUT queries, PALUTE leverages the vertical organization of M3D DRAM. Figure~\ref{fig:PALUTE_total}(c) illustrates the structure of an M3D DRAM bank, where each vertical stack corresponds to a DRAM MAT. Within each MAT, shown in Figure~\ref{fig:PALUTE_total}(d), every purple node represents a one transistor one cell (1T1C) DRAM cell that holds one bit of a stored LUT value. PALUTE exploits the vertical axis to issue LUT queries across many MATs simultaneously. This hardware-aware arrangement maximizes parallelism while keeping the area footprint minimal. Each MAT consists of horizontal WLs and vertical BLs, with a SA array located at the bottom of the stack. Beneath the SAs, transistor switches and FFs capture the selected LUT outputs. A MUX and a sign-flip unit implement the half-table LUT mechanism and convert raw table outputs into their corresponding signed values. On the left side of each MAT, matching logic integrated into a lightweight row decoder iteratively sweeps through all MAT rows, identifies matches with the queried index, and activates the corresponding group of switches that route the matched LUT columns to the FFs. As a MAT can store lots of LUTs horizontally, a single sweep retrieves LUT outputs from many LUTs in parallel, yielding high query throughput. Assume there are $M$ channels per chip, $N$ banks per channel and $K$ MATs per bank, reaching $M\times N\times K$ MATs for LUT query in parallel. Assume a row-buffer interface bandwidth of $BW_{buf}$ bit/cycle and the data storing in the MAT are $W$-bit, maximum parallelism reaches $M\times N\times K\times \frac{BW_{buf}}{W}$.

        \subsubsection{M3D DRAM Chip Architecture}
    
            A bank in our M3D DRAM architecture consists of 1024 MATs, and each channel is organized as a 4$\times$4 array of banks. Likewise, 4$\times$4 channels together form a full M3D DRAM chip, as illustrated in Figure~\ref{fig:PALUTE_total}(a). Each channel is interfaced with a LUT-generation unit on the logic die through hybrid bonding, which provides ultrahigh bandwidth communication (1.38 TB/s)~\cite{HybridBonding}. The paraellism can reach $8,388,608$ for W4A4 quantization and 128 bit/cycle~\cite{3D_Stacked_Memory_Bandwidth} maximum row-buffer interface bandwidth. All MATs in a bank are connected to a bank-level accumulator that aggregates partial results across matrix segments. Within each MAT, the storage space is logically partitioned into three functional regions: the upper region stores KV-cache data, the middle region stores LUTs and enables in-DRAM LUT querying, and the lower region stores model weights. This organization is motivated by the access-frequency and data-locality characteristics of each data type, whose structure is shown in Figure~\ref{fig:PALUTE_total}(b), as further discussed in the Sec.~\ref{Sec:Scheduling}. During computation, LUT-retrieved values are forwarded to the bank-level accumulator.

        \begin{figure}[ht]
            \centering
            \includegraphics[width=1\linewidth]{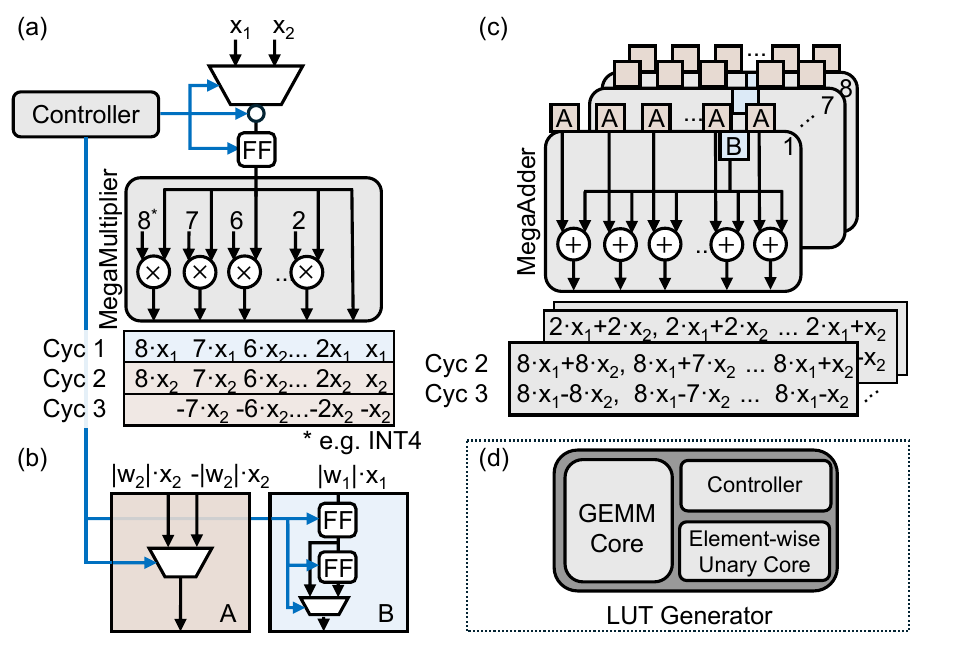}
            \caption{LUT generator (a) Digital logic for the first addend controlled by a FSM-based controller (b) Digital logic for state alignment (c) Digital logic for the second addend (d) Overall LUT generator architecture with GEMM and non-linear core}
            \label{fig:LUTGEN}
        \end{figure}

        \subsection{LUT Generator}\label{Sec:Hardware}
            The LUT generator is implemented on the logic die and connected to DRAM banks via hybrid bonding. It consists of a $4\times4$ array of units, where each unit serves one DRAM bank. As shown in Figure~\ref{fig:LUTGEN}(d), each unit integrates two specialized cores: a GEMM core for generating LUTs used in GEMM, and an element-wise unary core for generating LUTs for functions such as GELU. PALUTE’s bank-attached LUT generator is a system component that generates LUTs and writes them back into the M3D DRAM array for in-situ queries. PALUTE extends LUT generation beyond GEMM to also cover element-wise unary operators (e.g., GELU). Prior work~\cite{FIGLUT} generates LUTs on the fly and propagates them across PE arrays, which can repeatedly regenerate identical LUT contents with limited reuse and thus incurs recurrent latency and energy overhead. Prior works~\cite{PIMPAL,pLUTo,3D-PATH} assume that LUTs are precomputed and merely staged in memory, which fails to account for LUT generation overhead. PALUTE employs a bank-parallel LUT generator that materializes both GEMM and unary-operator LUTs directly near M3D DRAM and transfers them with high bandwidth.
            
            Instead of performing explicit multiply-accumulate (MAC) or dequantization operations during inference, the GEMM core enumerates all possible partial sums of low-precision activations and materializes them as LUT entries. Subsequent computation is reduced to index-based LUT lookup and accumulation, reducing arithmetic complexity and data movement. For a LUT entry $w_1x_1+w_2x_2$, we leverage the half-table by enforcing $w_1\!\ge\!0$ while allowing $w_2$ to be signed. The GEMM core therefore generates $|w_1|x_1$ and $\pm|w_2|x_2$ separately and combines them to enumerate all entries. Figure~\ref{fig:LUTGEN}(a) illustrates INT4 weights with $|w_1|\!\in\!\{1,\dots,8\}$ and $|w_2|\!\in\!\{1,\dots,7\}$. A controller drives a multiplier array plus a MUX with a sign-inverter to produce partial products in three cycles: (1) $|w_1|x_1$, (2) $|w_2|x_2$, and (3) $-|w_2|x_2$. In cycle~1, all $|w_1|x_1$ terms are buffered (Figure~\ref{fig:LUTGEN}(b) right); cycles~2--3 buffer the corresponding $\pm|w_2|x_2$ terms (Figure~\ref{fig:LUTGEN}(b) left). Cycle-aligned multiplexing pairs buffered $|w_1|x_1$ with each $\pm|w_2|x_2$ and forwards them to the adder arrays. Figure~\ref{fig:LUTGEN}(c) shows the replicated MegaAdder: each instance takes a fixed $|w_1|x_1$ on one input and enumerates all $\pm|w_2|x_2$ on the others, producing the full set of partial sums in parallel. Consequently, all LUT entries for $w_1x_1\pm w_2x_2$ are materialized within three cycles and then written back to DRAM.
            
        \subsection{PALUTE Tiering and Scheduling Framework}\label{Sec:Scheduling}
    
            Due to the staircase-like wordline (WL) routing in M3D DRAM, lower WLs suffer higher parasitic capacitance and resistance from longer routing paths. This latency imbalance becomes significant as M3D DRAM scales to hundreds of layers~\cite{Stratum}, leading to nonuniform activation latency across tiers and making data tiering necessary. Figure~\ref{fig:PALUTE_total}(b) illustrates PALUTE’s end-to-end scheduling flow: \encircle{1} weights and initial tokens are fetched into the logic die through hybrid bonding; \encircle{2} the LUT-generation unit constructs LUTs from incoming token values; \encircle{3} the LUTs are written back into the M3D DRAM array for in-situ queries; \encircle{4} weights serve as indices during LUT querying, and each MAT returns LUT results to the bank-level accumulator; \encircle{5} the KV cache is stored in M3D DRAM to support high-throughput decoding read/write.

            To reduce data movement, PALUTE places data according to tier locality. Initial tokens and frequently accessed weights are placed in lower tiers close to the logic die, reducing WL traversal length. LUTs are placed in middle tiers to balance latency and capacity. Since KV-cache read/write dominates decoding traffic, the KV cache is placed adjacent to LUT storage in higher layers. The bank-level accumulator sits at the top of each MAT, directly above the KV-cache region, so LUT results travel only a short vertical path before accumulation, reducing latency and energy while improving effective bandwidth.

\section{Evaluation}
        
    \subsection{Experimental Setup}
    \subsubsection{Baselines}
    
        We evaluate PALUTE against an NVIDIA Jetson Orin NX GPU~\cite{Jetson} and SOTA in-memory (normal DRAM) LUT-based accelerators PIMPAL~\cite{PIMPAL}, a LUT-based ASIC FIGLUT~\cite{FIGLUT} and a heterogeneous NMP accelerator CHIME~\cite{CHIME}. We assess end-to-end LLM inference, and we compare performance in terms of throughput, power, energy efficiency and area efficiency. All evaluations are conducted on the Qwen3-0.6B, 1.7B, 4B and 8B~\cite{Qwen3} model under W4A4 quantization.

        \subsubsection{Hardware Configuration}

            Table~\ref{tab:dram} summarizes the hardware configuration of the baseline M3D DRAM used in this work. We assume a 35\,nm M3D DRAM device with 768 vertically stacked layers. Each MAT is vertical and consists of a $768 \times 1024$ cell array. A total of 1024 MATs form one bank, and banks are hierarchically organized as a $4 \times 4$ array per channel and a $4 \times 4$ array of channels per chip, enabling massive parallelism across banks and channels.

            Under this organization, the effective storage capacity of a single layer is 32 MB, with 768 stacked layers, the total chip capacity reaches approximately 24 GB. We assume a row-buffer interface bandwidth of 128 bps~\cite{3D_Stacked_Memory_Bandwidth}, which determines the peak data transfer rate for row activations. Let each LUT result is stored in 4-bit, a 128-bit row-buffer burst can deliver at most 32 results per transfer, i.e., the peak transfer parallelism is 32. When more than 32 LUTs are activated, the results must be moved in multiple bursts, effectively serializing the transfer. The system sustains up to $16\times16\times1024\times32=8,388,608$ concurrent LUT lookups.
        
        \subsubsection{Methodology}
        
            We evaluate PALUTE with an in-house cycle-accurate simulator that models end-to-end LUT-based Transformer inference on M3D DRAM, including in-DRAM LUT querying, bank-level accumulation, and the near-memory LUT generator. The LUT-query timing follows pLUTo-style bulk LUT access semantics~\cite{pLUTo}, extended to M3D DRAM's vertical MAT organization. To model realistic overheads, we implement the LUT Generator and system controller are implemented in Verilog and synthesized with Cadence Genus using a 7\,nm CMOS PDK at 200\,MHz for performance, power, and area estimation.

            \begin{table}[t]
                \centering
                \caption{Hardware parameters of M3D DRAM}
                \label{tab:dram}
                \scalebox{0.85}{
                \renewcommand{\arraystretch}{1.1}
                \begin{tabular}{|l|c|l|c|}
                    \hline
                    \multicolumn{4}{|c|}{\textbf{Device Overview}} \\ 
                    \hline
                    \# Layers              & 768      & Technology Node & 35 nm \\
                    \hline
                    Bank Capacity          & 96 MB    & Bank Area      & 0.44 mm$^{2}$ \\
                    \hline
                    Chip Capacity          & 24 GB    & Chip Area      & 112.4 mm$^{2}$ \\
                    \hline
                    \multicolumn{4}{|c|}{\textbf{Energy Parameters (pJ / row)}} \\ 
                    \hline
                    $E_{\mathrm{ACT}}$ & 207  & $E_{\mathrm{PRE}}$ & 458 \\
                    \hline
                    $E_{\mathrm{READ}}$    & 7260   &  $E_{\mathrm{WRITE}}$ & 7540 \\
                    \hline
                    \multicolumn{4}{|c|}{\textbf{Timing Parameters (ns)}} \\ 
                    \hline
                    $T_{\mathrm{ACT}}$  & 14.16 & $T_{\mathrm{PRE}}$   & 14.16 \\
                    \hline
                    $T_{\mathrm{READ}}$ & 14.16 & $T_{\mathrm{WRITE}}$ & 14.16 \\
                    \hline
                \end{tabular}
                }
            \end{table}

        \begin{table}[t]
            \centering
            \small
            \caption{End-to-end energy and area efficiency comparison}
            \label{tab:end2end_eff}
            \begin{tabular}{lcccc}
                \hline
                \textbf{} & \textbf{PIMPAL} & \textbf{CHIME} & \textbf{FIGLUT} & \textbf{Ours} \\
                \hline
                Category               & PIM   & NMP    & ASIC  & PIM     \\
                Technology$^{*}$ (nm)  & N/A   & 7nm    & 28nm  & 7nm     \\
                Power (W)              & N/A   & 1.95   & 0.29  & 0.16    \\
                Area$^{**}$ (mm$^2$)   & 1.82 & 53.6  & 0.78  & 112.4  \\
                Throughput (TPS)       & 10.2 & 1,179   & 20.7  & 1,264 \\
                Energy Eff. (TPS/W)    & N/A   & 604.6 & 71.4 & 7,738 \\
                Area Eff. (TPS/mm$^2$) & 5.62  & 22.01  & 26.54 & 11.25   \\
                \hline
            \end{tabular}
        
            \footnotesize
            \raggedright
            \noindent $^{*}$ For PIM/NMP baselines, the technology node is for the logic die. \\
            \noindent $^{**}$ For PIM/NMP baselines, the area is for the max\{memory, logic die\}.
        \end{table}
    % Figure~\ref{fig:Evaluation_GEMM},~\ref{fig:Evaluation_Jetson} and Table~\ref{tab:end2end_eff} jointly compare PALUTE against SOTA accelerators PIMPAL~\cite{PIMPAL}, FIGLUT~\cite{FIGLUT}, CHIME~\cite{CHIME} and Jetson Orin NX~\cite{Jetson} in terms of throughput, power, energy efficiency, and area efficiency under W4A4 precision.
    
        % \begin{figure}[ht]
        %     \centering
        %     \includegraphics[width=1\linewidth]{figures/Evaluation_GEMM.pdf}
        %     \caption{GEMM energy and area efficiency comparison}
        %     \label{fig:Evaluation_GEMM}
        % \end{figure}

        % We first benchmark the GEMM operator. Figure~\ref{fig:Evaluation_GEMM} shows energy efficiency and area efficiency across four matrix sizes (1k$\times$1k to 8k$\times$8k). We report equivalent TOPS by mapping LUT-based compute to the standard GEMM MAC operation count (one multiply and one add are counted as two operations). PALUTE consistently achieves the best energy efficiency across all sizes (Figure~\ref{fig:Evaluation_GEMM}(a)), outperforming PIMPAL by average 1895$\times$ and FIGLUT by 5563$\times$. In contrast, PALUTE performs LUT-query-based computation near/in memory and avoids dequantization, thereby improving effective compute utilization and reducing energy cost per operation.

    \subsection{Performance}

        % Table~\ref{tab:end2end_eff} and Figure~\ref{fig:Evaluation_Jetson} jointly compare PALUTE against SOTA accelerators~\cite{PIMPAL, FIGLUT, CHIME} and Jetson Orin NX~\cite{Jetson}. Table~\ref{tab:end2end_eff} summarizes the end-to-end energy and area efficiency of PALUTE for full-model decoding under W4A4 precision. PALUTE achieves an energy efficiency of 7,738, outperforming CHIME~\cite{CHIME} and FIGLUT~\cite{FIGLUT} by 12.8$\times$ and 1.6$\times$, respectively; PIMPAL~\cite{PIMPAL} does not report end-to-end energy in its evaluation. For area efficiency, PALUTE reaches 11.25~TPS/mm$^2$, exceeding PIMPAL~\cite{PIMPAL} by 2.0$\times$. We restrict the area-efficiency comparison to in-memory LUT-based designs, since LUT support introduces non-trivial area overhead that is explicitly accounted for in our implementation but is absent in non-LUT baselines. Figure~\ref{fig:Evaluation_Jetson}(a) further compares PALUTE against Jetson~\cite{Jetson}, showing consistent gains in both energy and area efficiency across model sizes from 0.6B to 8B. Figure~\ref{fig:Evaluation_Jetson}(b) places PALUTE on a favorable throughput--power Pareto frontier relative to Jetson~\cite{Jetson}, FIGLUT~\cite{FIGLUT}, and CHIME~\cite{CHIME}, demonstrating higher token throughput under a tighter power envelope.

\begin{figure}[t] 
            \centering 
            \includegraphics[width=1\linewidth]{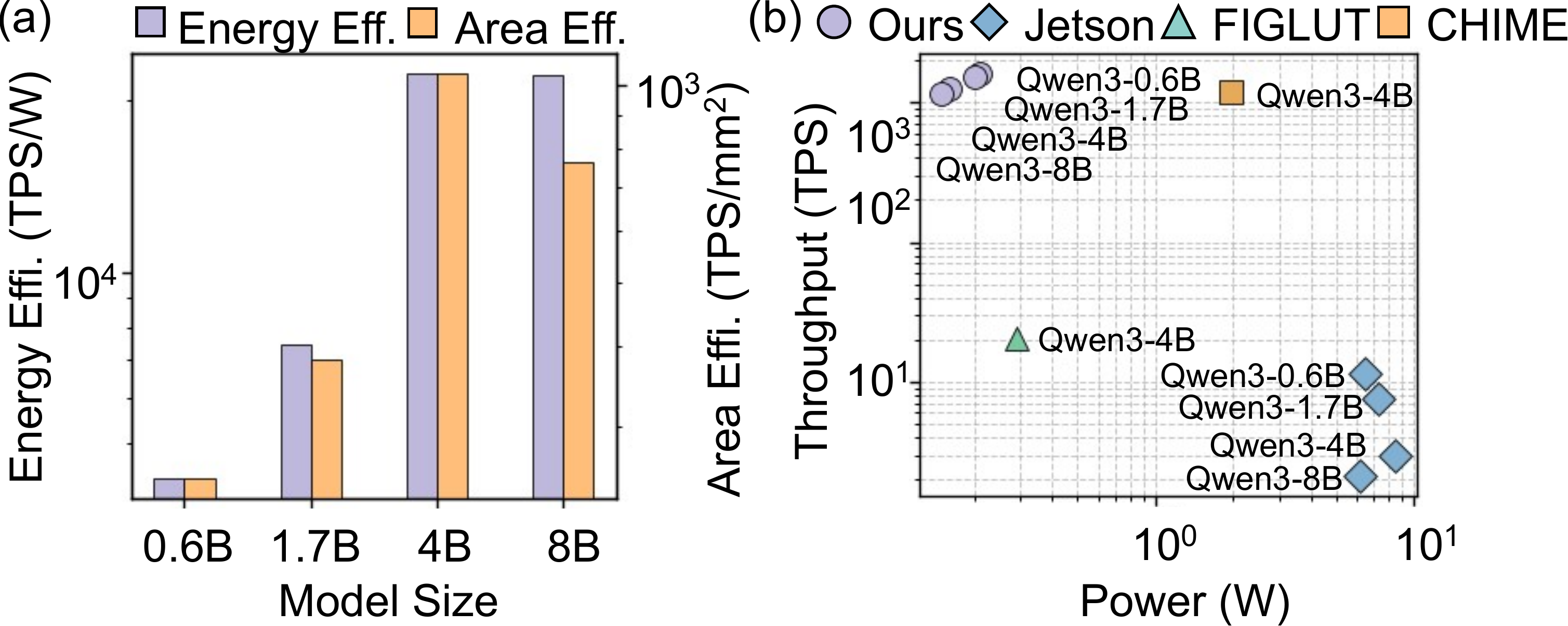} 
            \caption{End-to-end inference evaluation (a) Energy and area efficiency gain compared with Jetson (b) Throughput and power compared with all SOTAs} 
            \label{fig:Evaluation_Jetson} 
        \end{figure}

        Table~\ref{tab:end2end_eff} and Figure~\ref{fig:Evaluation_Jetson} jointly compare PALUTE against SOTA accelerators~\cite{PIMPAL, FIGLUT, CHIME} and Jetson Orin NX~\cite{Jetson}. Table~\ref{tab:end2end_eff} summarizes the end-to-end energy and area efficiency of PALUTE for decoding under W4A4 precision. PALUTE achieves an energy efficiency of 7,738, outperforming CHIME~\cite{CHIME} and FIGLUT~\cite{FIGLUT} by 12.8$\times$ and 1.6$\times$, respectively; PIMPAL~\cite{PIMPAL} does not report end-to-end energy in its evaluation. These gains come from reducing the energy from dequantization, GEMM and nonlinear kernel, as well as the system-level data movement. For area efficiency, PALUTE reaches 11.25~TPS/mm$^2$, exceeding PIMPAL~\cite{PIMPAL} by 2.0$\times$, which is driven by M3D DRAM's vertical MAT organization. We restrict the area-efficiency comparison to in-memory LUT-based designs, since LUT support introduces non-trivial area overhead that is explicitly accounted for in our implementation but is absent in non-LUT baselines. Figure~\ref{fig:Evaluation_Jetson}(a) compares PALUTE against Jetson~\cite{Jetson}, showing consistent gains in both energy and area efficiency across model sizes from 0.6B to 8B. Figure~\ref{fig:Evaluation_Jetson}(b) places PALUTE on a favorable throughput--power Pareto frontier relative to Jetson~\cite{Jetson}, FIGLUT~\cite{FIGLUT}, and CHIME~\cite{CHIME}, demonstrating higher token throughput under a tighter power envelope.
        
        \begin{figure}[t]
            \centering
            \includegraphics[width=1\linewidth]{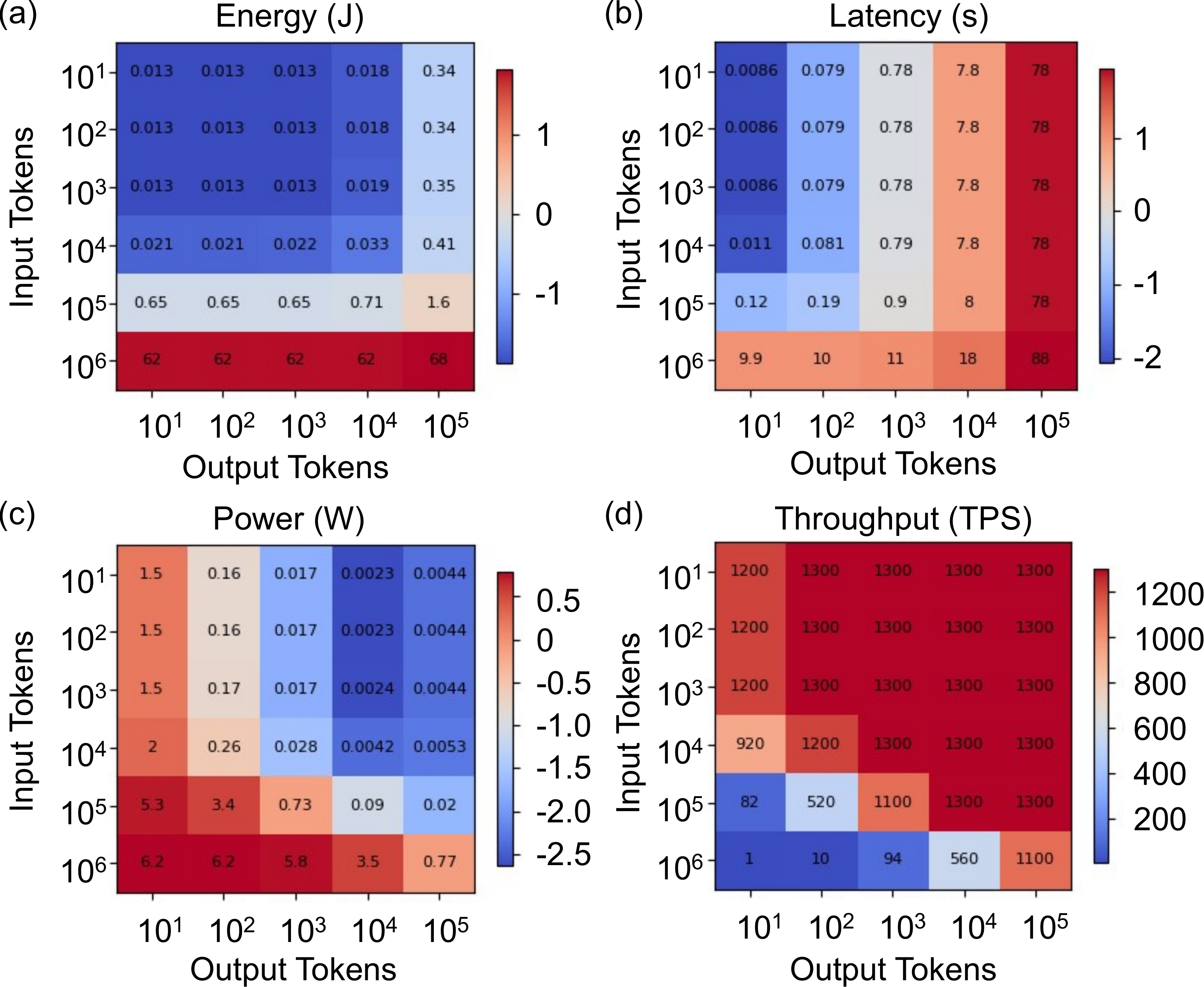}
            \caption{Sensitivity to input/output token lengths: heatmaps of (a) energy, (b) latency, (c) power, and (d) throughput}
            \label{fig:heatmap}
        \end{figure}

        \subsection{Sensitivity Analysis}

        This section studies PALUTE's sensitivity to prompt (input) and generation (output) lengths by sweeping token counts and reporting energy and latency. As Figure~\ref{fig:heatmap}(a) and (b) shows, both metrics generally rise with longer sequences because longer decoding increases auto-regressive steps and enlarges the KV cache, raising memory traffic and attention cost.

        For a fixed output length, Figure~\ref{fig:heatmap}(c) shows that shorter inputs can reduce average power by decreasing prefill work and the initial KV cache size; decoding power scales largely with the active KV footprint. Figure~\ref{fig:heatmap}(d) shows the corresponding throughput: shorter inputs and longer outputs typically improve throughput since prefill overhead is amortized and KV reuse is higher. In contrast, long inputs spend more time in prefill and create a large KV cache early, increasing per-step cost and lowering steady-state token rate; very short outputs also limit KV reuse and reduce effective throughput.

\begin{figure}[t]
            \centering
            \includegraphics[width=0.95\linewidth]{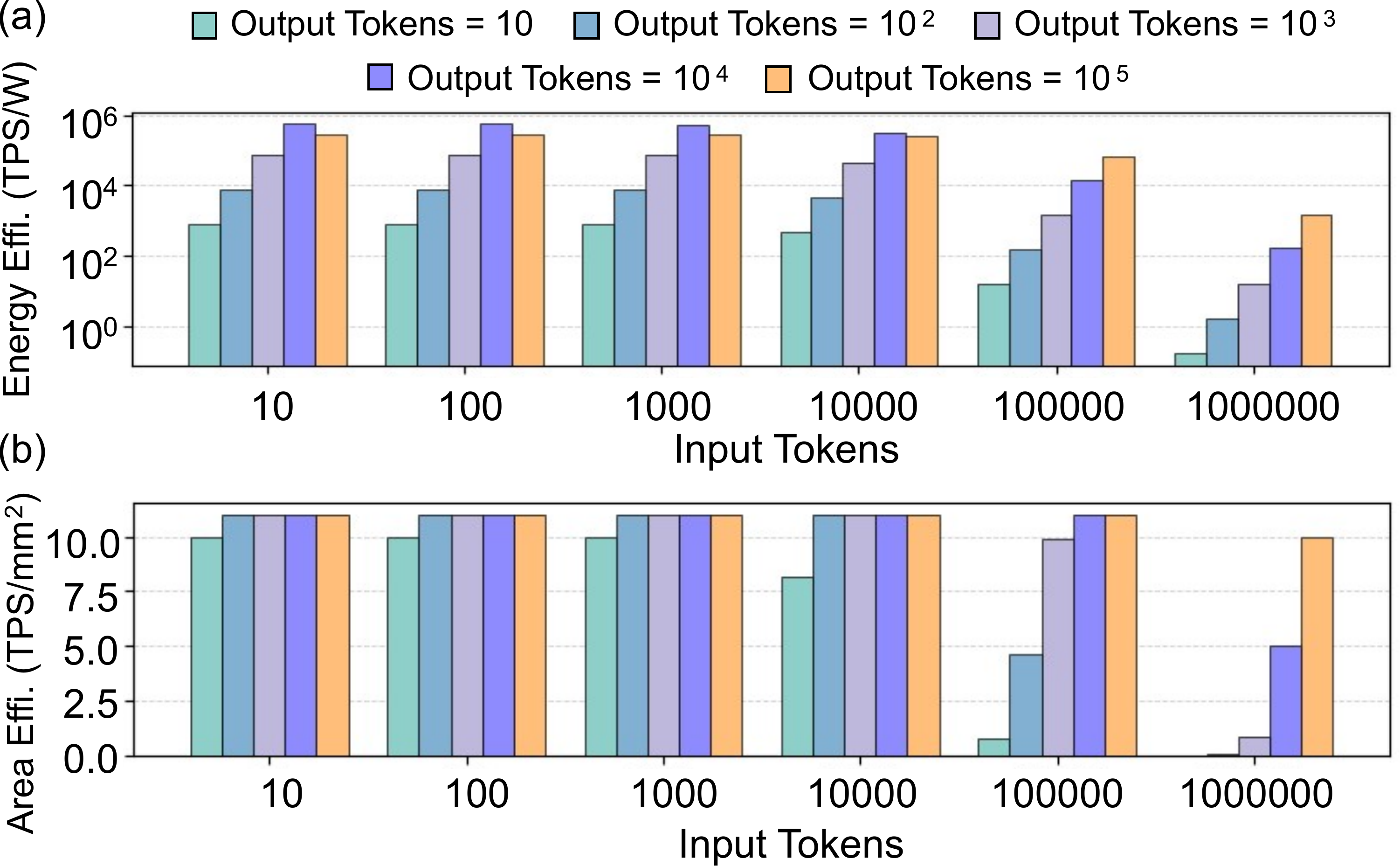}
            \caption{Energy and area efficiency versus input token length under different output-length settings (a) Energy (b) Area}
            \label{fig:sweep_token_len}
        \end{figure}

        % We further examine how energy efficiency and area efficiency vary with sequence length. Figure~\ref{fig:sweep_token_len}(a) shows that increasing input length tends to reduce energy efficiency, whereas increasing output length generally improves energy efficiency, consistent with the power/throughput behaviors above. However, when the output length becomes sufficiently large under long-input settings, the throughput degradation induced by the large KV cache can dominate, offsetting the amortization benefit and leading to a lower net energy efficiency. For area efficiency, we compute \textit{AreaEff} as $\mathrm{Throughput}/\mathrm{Area}$. Since PALUTE’s silicon area is fixed in this sweep, Figure~\ref{fig:sweep_token_len}(b) directly mirrors the throughput trend in Figure~\ref{fig:heatmap}(d): configurations that achieve higher throughput also achieve higher area efficiency, and vice versa.
        We further evaluate energy and area efficiency versus sequence length. Figure~\ref{fig:sweep_token_len}(a) shows energy efficiency decreases with longer inputs but generally improves with longer outputs, unless a large KV cache under long-input settings degrades throughput enough to offset the benefit of longer decoding. We define $\mathrm{AreaEff}=\mathrm{Throughput}/\mathrm{Area}$; since area is constant, Figure~\ref{fig:sweep_token_len}(b) directly follows the throughput trend in Figure~\ref{fig:heatmap}(d).
        
        \begin{figure}[t]
            \centering
            \includegraphics[width=0.95\linewidth]{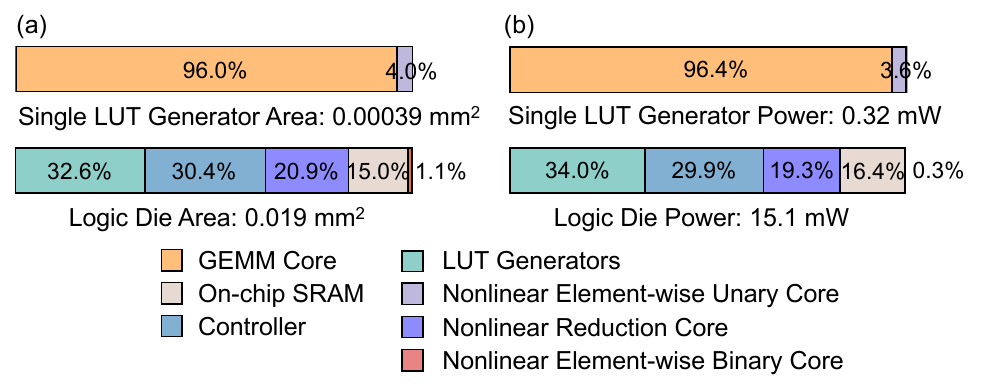}
            \caption{Overhead breakdown (a) Area (b) Power}
            \label{fig:overhead}
        \end{figure}

        \subsection{Overhead Breakdown and Analysis}
        
        On the logic die, PALUTE integrates a controller and LUT generators. For a single LUT generator (Fig.~\ref{fig:overhead}(a) and (b)), the GEMM core dominates both area and power, accounting for 96.0\% of 0.00039\,mm$^2$ and over 96.4\% of 0.32\,mW, while the nonlinear element-wise unary core contributes only 4.0\% area and 2.06\% power.

        At the logic-die level, LUT generators occupy 32.6\% of the logic-die area, followed by the controller (30.4\%), nonlinear reduction core (20.9\%), and SRAM (15.0\%), with the nonlinear element-wise binary core negligible (1.1\%). Total logic-die area is 0.019\,mm$^2$. Power is led by LUT-related computation (34.0\%), followed by the controller (29.9\%), nonlinear reduction core (19.3\%) and On-chip SRAM (16.4\%). Overall logic-die power is 15.1\,mW.
        
\section{Conclusion}

       We present PALUTE, a LUT-based PIM accelerator on M3D DRAM that leverages in-DRAM LUT querying, a logic-die LUT generator, and system-level tiering/scheduling to cut dequantization, nonlinear overheads, and data movement for edge LLM inference. Under end-to-end W4A4 quantization, PALUTE delivers 1,264 TPS at 0.16 W (7,738 TPS/W), improving energy efficiency by 12.8$\times$ over CHIME~\cite{CHIME} and 1.6$\times$ over FIGLUT~\cite{FIGLUT}, and area efficiency by 2.0$\times$ over PIMPAL~\cite{PIMPAL}, demonstrating a practical path to energy/area-efficient LLM inference via LUT-centric execution on M3D DRAM.

\section*{Acknowledgment}
This work was supported in part by PRISM and CoCoSys, centers in JUMP 2.0, an SRC program sponsored by DARPA (SRC grant number 2023-JU-3135). This work was also supported by NSF grants \#2003279, \#1911095, \#2112167, \#2052809, \#2112665, \#2120019, \#2211386.

\bibliographystyle{ACM-Reference-Format}
\bibliography{sample-base}

@article{Mobile_Edge_Survey,
author = {Qu, Guanqiao and Chen, Qiyuan and Wei, Wei and Lin, Zheng and Chen, Xianhao and Huang, Kaibin},
year = {2025},
month = {01},
pages = {1-1},
title = {Mobile Edge Intelligence for Large Language Models: A Contemporary Survey},
volume = {PP},
journal = {IEEE Communications Surveys \& Tutorials},
doi = {10.1109/COMST.2025.3527641}
}

@misc{LLM_Survey,
	title = {A {Survey} of {Large} {Language} {Models}},
	url = {http://arxiv.org/abs/2303.18223},
	language = {en},
	urldate = {2023-09-14},
	publisher = {arXiv},
	author = {Zhao, Wayne Xin and Zhou, Kun and Li, Junyi and Tang, Tianyi and Wang, Xiaolei and Hou, Yupeng and Min, Yingqian and Zhang, Beichen and Zhang, Junjie and Dong, Zican and Du, Yifan and Yang, Chen and Chen, Yushuo and Chen, Zhipeng and Jiang, Jinhao and Ren, Ruiyang and Li, Yifan and Tang, Xinyu and Liu, Zikang and Liu, Peiyu and Nie, Jian-Yun and Wen, Ji-Rong},
	month = sep,
	year = {2023},
	note = {arXiv:2303.18223 [cs]},
}

@unknown{Healthcare_Agent,
author = {Tu, Tao and Azizi, Shekoofeh and Driess, Danny and Schaekermann, Mike and Amin, Mohamed and Chang, Pi-Chuan and Carroll, Andrew and Lau, Chuck and Tanno, Ryutaro and Ktena, Sofia Ira and Mustafa, Basil and Chowdhery, Aakanksha and Liu, Yun and Kornblith, Simon and Fleet, David and Mansfield, Philip and Prakash, Sushant and Wong, Renee and Virmani, Sunny and Natarajan, Vivek},
year = {2023},
month = {07},
pages = {},
title = {Towards Generalist Biomedical AI},
doi = {10.48550/arXiv.2307.14334}
}

@inproceedings{Intelligent_systems,
author = {Driess, Danny and Xia, Fei and Sajjadi, Mehdi S. M. and Lynch, Corey and Chowdhery, Aakanksha and Ichter, Brian and Wahid, Ayzaan and Tompson, Jonathan and Vuong, Quan and Yu, Tianhe and Huang, Wenlong and Chebotar, Yevgen and Sermanet, Pierre and Duckworth, Daniel and Levine, Sergey and Vanhoucke, Vincent and Hausman, Karol and Toussaint, Marc and Greff, Klaus and Zeng, Andy and Mordatch, Igor and Florence, Pete},
title = {PaLM-E: an embodied multimodal language model},
year = {2023},
publisher = {JMLR.org},
booktitle = {Proceedings of the 40th International Conference on Machine Learning},
articleno = {340},
numpages = {20},
location = {Honolulu, Hawaii, USA},
series = {ICML'23}
}

@misc{Quantization_Survey,
      title={A Survey of Quantization Methods for Efficient Neural Network Inference}, 
      author={Amir Gholami and Sehoon Kim and Zhen Dong and Zhewei Yao and Michael W. Mahoney and Kurt Keutzer},
      year={2021},
      eprint={2103.13630},
      archivePrefix={arXiv},
      primaryClass={cs.CV},
      url={https://arxiv.org/abs/2103.13630}, 
}

@misc{QServe,
      title={QServe: W4A8KV4 Quantization and System Co-design for Efficient LLM Serving}, 
      author={Yujun Lin and Haotian Tang and Shang Yang and Zhekai Zhang and Guangxuan Xiao and Chuang Gan and Song Han},
      year={2025},
      eprint={2405.04532},
      archivePrefix={arXiv},
      primaryClass={cs.CL},
      url={https://arxiv.org/abs/2405.04532}, 
}

@article{PIM_Survey,
author = {Asifuzzaman, Kazi and Miniskar, Narasinga and Young, Aaron and Liu, Frank and Vetter, Jeffrey},
year = {2022},
month = {12},
pages = {100022},
title = {A survey on processing-in-memory techniques: Advances and challenges},
volume = {4},
journal = {Memories - Materials, Devices, Circuits and Systems},
doi = {10.1016/j.memori.2022.100022}
}

@inproceedings{pLUTo,
  author    = {Ferreira, Joao Dinis and Falcao, Gabriel and Gomez-Luna, Juan and Alser, Mohammed and Orosa, Lois and Sadrosadati, Mohammad and Kim, Jeremie S. and Oliveira, Geraldo F. and Shahroodi, Taha and Nori, Anant and Mutlu, Onur},
  title     = {pLUTo: Enabling Massively Parallel Computation in DRAM via Lookup Tables},
  booktitle = {2022 IEEE/ACM International Symposium on Microarchitecture (MICRO)},
  year      = {2022},
  pages     = {900--919}
}

@INPROCEEDINGS{FIGLUT,
  author    = {Gunho Park and Hyeokjun Kwon and Jiwoo Kim and Jeongin Bae and Baeseong Park and Dongsoo Lee and Youngjoo Lee},
  title     = {FIGLUT: An Energy-Efficient Accelerator Design for FP-INT GEMM Using Look-Up Tables},
  booktitle = {IEEE International Symposium on High Performance Computer Architecture (HPCA)},
  year      = {2025},
  pages     = {1098--1111}
}

@inproceedings{Data_Movement_Bottlenecks,
  author    = {Boroumand, Amirali and Ghose, Saugata and Kim, Youngsok and Ausavarungnirun, Rachata and Shiu, Eric and Thakur, Rahul and Kim, Daehyun and Kuusela, Aki and Knies, Allan and Ranganathan, Parthasarathy and Mutlu, Onur},
  title     = {Google Workloads for Consumer Devices: Mitigating Data Movement Bottlenecks},
  booktitle = {Proceedings of the Twenty-Third International Conference on Architectural Support for Programming Languages and Operating Systems (ASPLOS)},
  year      = {2018},
  pages     = {316--331}
}

@inproceedings{Atom,
  author    = {Yilong Zhao and Chien-Yu Lin and Kan Zhu and Zihao Ye and Lequn Chen and Size Zheng and Luis Ceze and Arvind Krishnamurthy and Tianqi Chen and Baris Kasikci},
  title     = {Atom: Low-Bit Quantization for Efficient and Accurate LLM Serving},
  booktitle = {Proceedings of Machine Learning and Systems (MLSys)},
  year      = {2024},
  pages     = {196--209},
  volume    = {6}
}

@inproceedings{QuaRot,
  author    = {Saleh Ashkboos and Amirkeivan Mohtashami and Maximilian L. Croci and Bo Li and Pashmina Cameron and Martin Jaggi and Dan Alistarh and Torsten Hoefler and James Hensman},
  title     = {QuaRot: outlier-free 4-bit inference in rotated LLMs},
  booktitle = {Proceedings of the 38th International Conference on Neural Information Processing Systems (NeurIPS)},
  year      = {2024}
}

@article{I-BERT,
  author  = {Sehoon Kim and Amir Gholami and Zhewei Yao and Michael W. Mahoney and Kurt Keutzer},
  title   = {I-BERT: Integer-only BERT Quantization},
  journal = {International Conference on Machine Learning (ICML)},
  year    = {2021}
}

@inproceedings{3D-PATH,
  author    = {Yue, Zhiheng and Wang, Yang and Li, Chao and Wei, Shaojun and Hu, Yang and Yin, Shouyi},
  title     = {3D-PATH: A Hierarchy LUT Processing-in-memory Accelerator with Thermal-aware Hybrid Bonding Integration},
  booktitle = {Proceedings of the 58th IEEE/ACM International Symposium on Microarchitecture (MICRO)},
  year      = {2025},
  pages     = {78--93}
}

@INPROCEEDINGS{HybridBonding,
  author    = {Niu, Dimin and Li, Shuangchen and Wang, Yuhao and Han, Wei and Zhang, Zhe and Guan, Yijin and Guan, Tianchan and Sun, Fei and Xue, Fei and Duan, Lide and Fang, Yuanwei and Zheng, Hongzhong and Jiang, Xiping and Wang, Song and Zuo, Fengguo and Wang, Yubing and Yu, Bing and Ren, Qiwei and Xie, Yuan},
  title     = {184QPS/W 64Mb/mm2 3D Logic-to-DRAM Hybrid Bonding with Process-Near-Memory Engine for Recommendation System},
  booktitle = {IEEE International Solid-State Circuits Conference (ISSCC)},
  year      = {2022},
  pages     = {1--3}
}

@inproceedings{Stratum,
  author    = {Pan, Yue and Xia, Zihan and Hsu, Po-Kai and Hu, Lanxiang and Kim, Hyungyo and Sharda, Janak and Zhou, Minxuan and Kim, Nam Sung and Yu, Shimeng and Rosing, Tajana and Kang, Mingu},
  title     = {Stratum: System-Hardware Co-Design with Tiered Monolithic 3D-Stackable DRAM for Efficient MoE Serving},
  booktitle = {Proceedings of the 58th IEEE/ACM International Symposium on Microarchitecture (MICRO)},
  year      = {2025},
  pages     = {1--17}
}

@inproceedings{Enabling_In-Memory_Computation,
  author    = {Onur Mutlu and Saugata Ghose and Juan Gómez-Luna and Rachata Ausavarungnirun},
  title     = {Processing Data Where It Makes Sense in Modern Computing Systems: Enabling In-Memory Computation},
  booktitle = {Great Lakes Symposium on VLSI (GLSVLSI)},
  year      = {2019},
  pages     = {5--6}
}

@inproceedings{Transformer_Breakdown,
  author    = {Sehoon Kim and Coleman Hooper and Thanakul Wattanawong and Minwoo Kang and Ruohan Yan and Hasan Genc and Grace Dinh and Qijing Huang and Kurt Keutzer and Michael W. Mahoney and Yakun Sophia Shao and Amir Gholami},
  title     = {Full Stack Optimization of Transformer Inference},
  booktitle = {Architecture and System Support for Transformer Models (ASSYST @ ISCA)},
  year      = {2023}
}

@misc{DATA_MOVEMENT_IS_ALL_YOU_NEED,
  author        = {Andrei Ivanov and Nikoli Dryden and Tal Ben-Nun and Shigang Li and Torsten Hoefler},
  title         = {Data Movement Is All You Need: A Case Study on Optimizing Transformers},
  year          = {2021},
  archivePrefix = {arXiv},
  eprint        = {2007.00072},
  primaryClass  = {cs.LG}
}

@inproceedings{FLASHATTENTION,
  author    = {Tri Dao and Daniel Y. Fu and Stefano Ermon and Atri Rudra and Christopher Ré},
  title     = {FLASHATTENTION: fast and memory-efficient exact attention with IO-awareness},
  booktitle = {Advances in Neural Information Processing Systems (NeurIPS)},
  year      = {2022}
}

@ARTICLE{M3DDRAM,
  author  = {Hsu, Po-Kai and Sharda, Janak and Wu, Xiangjin and Wong, H.-S. Philip and Yu, Shimeng},
  title   = {Monolithic 3D Stackable DRAM},
  journal = {IEEE Nanotechnology Magazine},
  year    = {2025},
  volume  = {19},
  number  = {2},
  pages   = {7--16}
}

@techreport{Jetson,
  author = {{NVIDIA Corporation}},
  title  = {Jetson Orin NX Series Datasheet},
  year   = {2022}
}

@misc{Qwen3,
  author        = {An Yang and Anfeng Li and Baosong Yang and Beichen Zhang and Binyuan Hui and Bo Zheng and Bowen Yu and Chang Gao and Chengen Huang and Chenxu Lv and Chujie Zheng and Dayiheng Liu and Fan Zhou and Fei Huang and Feng Hu and Hao Ge and Haoran Wei and Huan Lin and Jialong Tang and Jian Yang and Jianhong Tu and Jianwei Zhang and Jianxin Yang and Jiaxi Yang and Jing Zhou and Jingren Zhou and Junyang Lin and Kai Dang and Keqin Bao and Kexin Yang and Le Yu and Lianghao Deng and Mei Li and Mingfeng Xue and Mingze Li and Pei Zhang and Peng Wang and Qin Zhu and Rui Men and Ruize Gao and Shixuan Liu and Shuang Luo and Tianhao Li and Tianyi Tang and Wenbiao Yin and Xingzhang Ren and Xinyu Wang and Xinyu Zhang and Xuancheng Ren and Yang Fan and Yang Su and Yichang Zhang and Yinger Zhang and Yu Wan and Yuqiong Liu and Zekun Wang and Zeyu Cui and Zhenru Zhang and Zhipeng Zhou and Zihan Qiu},
  title         = {Qwen3 Technical Report},
  year          = {2025},
  archivePrefix = {arXiv},
  eprint        = {2505.09388},
  primaryClass  = {cs.CL}
}

@INPROCEEDINGS{PIMPAL,
  author    = {Jang, Yoonho and Cho, Hyeongjun and Ryu, Yesin and Kim, Jungrae and Hong, Seokin},
  title     = {PIMPAL: Accelerating LLM Inference on Edge Devices via In-DRAM Arithmetic Lookup},
  booktitle = {Proceedings of the 62nd Annual ACM/IEEE Design Automation Conference (DAC)},
  year      = {2025}
}

@misc{CHIME,
  author        = {Yanru Chen and Runyang Tian and Yue Pan and Zheyu Li and Weihong Xu and Tajana Rosing},
  title         = {CHIME: Chiplet-based Heterogeneous Near-Memory Acceleration for Edge Multimodal LLM Inference},
  year          = {2025},
  archivePrefix = {arXiv},
  eprint        = {2601.19908},
  primaryClass  = {cs.AR}
}

@article{3D_Stacked_Memory_Bandwidth,
  author  = {Donghyuk Lee and Gennady Pekhimenko and Samira Khan and Saugata Ghose and Onur Mutlu},
  title   = {Simultaneous Multi-Layer Access: Improving 3D-Stacked Memory Bandwidth at Low Cost},
  journal = {ACM Transactions on Architecture and Code Optimization},
  year    = {2016},
  volume  = {12},
  number  = {4}
}

@inproceedings{TMAC,
author = {Wei, Jianyu and Cao, Shijie and Cao, Ting and Ma, Lingxiao and Wang, Lei and Zhang, Yanyong and Yang, Mao},
title = {T-MAC: CPU Renaissance via Table Lookup for Low-Bit LLM Deployment on Edge},
year = {2025},
isbn = {9798400711961},
publisher = {Association for Computing Machinery},
address = {New York, NY, USA},
booktitle = {Proceedings of the Twentieth European Conference on Computer Systems},
pages = {278–292},
numpages = {15},
location = {Rotterdam, Netherlands},
series = {EuroSys '25}
}

\end{document}